\documentclass[twocolumn,showpacs,preprintnumbers,prl]{revtex4}
\usepackage{graphicx}
\usepackage{dcolumn}
\usepackage{bm}
\usepackage{color}

\begin{document}

\title{Bright Single Photons for Light Matter Interaction}
\author{Chih-Hsiang Wu}
\author{Tsung-Yao Wu\footnote[2]{Present address: Department of Physics, The Pennsylvania State University, USA.}}
\author{Yung-Chin Yeh}
\author{Po-Hui Liu}
\author{Chin-Hsuan Chang}
\author{Chiao-Kai Liu}
\author{Ting Cheng}
\author{Chih-Sung Chuu}
\email{cschuu@phys.nthu.edu.tw}
\affiliation{Department of Physics, National Tsing Hua University, Hsinchu 30013, Taiwan\\
and Frontier Research Center on Fundamental and Applied Sciences of Matters, National Tsing Hua University, Hsinchu 30013, Taiwan}

\begin{abstract}  
Single photons of subnatural linewidth and high spectral brightness are necessary for efficient light-matter interaction at the single photon level, which lies at the heart of many quantum photonic technologies. Here we demonstrate a bright source of single photons with subnatural linewidth, controllable waveforms, and a high spectral brightness of $3.67 \times 10^5$~s$^{-1}$~mW$^{-1}$~MHz$^{-1}$. The interaction between the single photons and atoms is demonstrated by the controlled absorption of the single photons in an atomic vapor. Our work has potential applications in quantum information technologies.
\end{abstract}


\maketitle

\section{I. Introduction}

A quantum network \cite{Kimble08} provides opportunities for advanced photonic technologies such as quantum computation, quantum communication, and quantum metrology. In a quantum network, quantum information is transmitted by single photons and processed or stored at quantum nodes. To implement the quantum nodes, the atomic ensemble has emerged as a promising physical system over recent years \cite{Duan01, Lukin03, Sangouard11}. Efficient interaction between single photons and atoms is thus crucial to successfully map the quantum states from one node to another in a quantum network. Such light-matter interaction necessitates single photons of subnatural linewidth and high spectral brightness. 

So far approaches of four-wave mixing with electromagnetically induced transparency \cite{Balic05, Du08}, resonant parametric down-conversion \cite{Kuklewicz06, Bao08, Scholz09, Wolfgramm11, Chuu12, Luo15}, and cavity quantum electrodynamics \cite{Kuhn02, Keller04, McKeever04, Thompson06} have been explored to generate single photons for light-matter interaction at the single-photon level. However, none of the single-photon sources has yet fulfilled a subnatural linewidth and high spectral brightness at the same time. Moreover, the highest generated spectral brightness, namely, the generated photon rate per MHz of bandwidth and per mW of pump power, reported to date is only $3 \times 10^4$~s$^{-1}$~MHz$^{-1}$~mW$^{-1}$~\cite{Luo15}. 

In this paper we demonstrate a subnatural-linewidth single-photon source with a spectral brightness one order of magnitude higher than previously reported. The interaction between the single photons and atoms is demonstrated by the controlled absorption of the single photons in a rubidium vapor. In addition, the long temporal width of the single photons allows us to control the waveform of the single photons from spontaneous parametric down-conversion. Our work thus also finds applications such as optimum quantum state mapping in a quantum network~\cite{Cirac97}, coherent control of storage, retrieval, and absorption of single photons in atomic ensembles \cite{Gorshkov07, Novikova07, Zhang12}, and quantum key distribution with high key creation efficiency \cite{Inoue02, Liu13}, in which waveform-controlled single photons are required. 

\section{II. SINGLE-MODE SINGLE PHOTONS}

The generation of our single photons is based on the resonant parametric down-conversion, but with several vital features added to ensure its ultrahigh spectral brightness. These features include the doubling of the interaction length between the pump, signal, and idler photons, the creation of the mode clusters at the signal and idler frequencies, and the maximization of the mode spacing. Every feature is indispensable for our single-photon source to operate in a single mode without the need of external filtering \cite{Chuu12, Chuu11}. This is the key to greatly increase the spectral brightness as compared to a multimode single-photon source, in which several tens of modes will be present if external filtering is not exploited. 

In order to obtain a single mode under the gain curve of the parametric down-conversion, the gain linewidth should be narrower than the mode spacing of the single photons. In our experiment, we minimize the gain linewidth by employing the double-pass pumping and type-II phase-matching, with a crucial factor of 2 provided by the former. The resulting gain linewidth is $0.44c/\Delta n_g L$, where $\Delta n_g$ is the differential group index of the signal and idler photons and $L$ is the crystal length. We also maximize the mode spacing by using a monolithic cavity and doubly resonating the signal and idler photons, of which the free spectral ranges are different due to orthogonal polarizations. This results in doubly resonant pairs of signal and idler modes [Fig.~\ref{fig:1}(a)], with a spacing $0.5c/\Delta n_g L$ larger than the gain linewidth of the parametric down-conversion. In addition, the monolithic cavity provides high mechanical stability and compactness for our single photon source.

\section{III. EXPERIMENTS}

\begin{figure}[t]
\centering
\includegraphics[width=0.9 \linewidth]{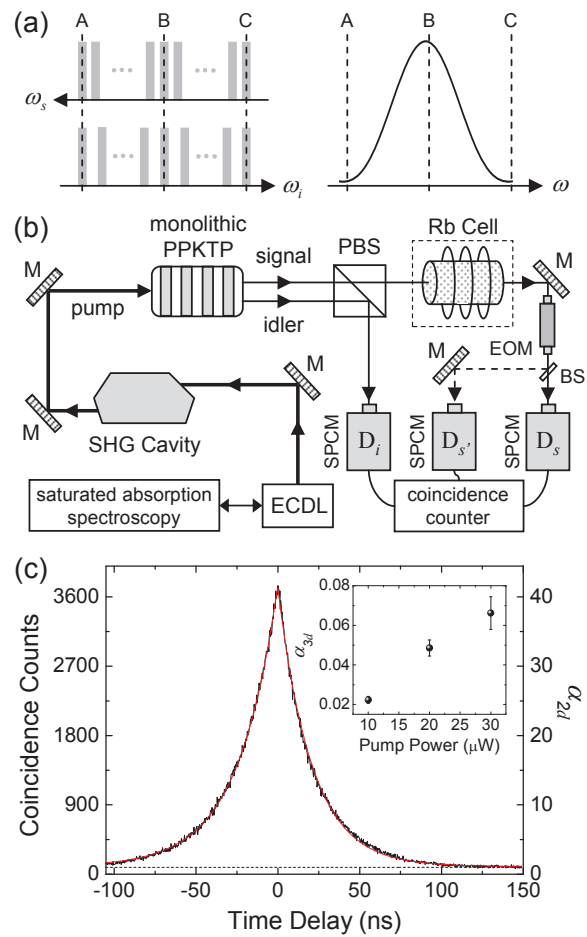}
\caption{\label{fig:1} (a) The left panel shows the signal and idler modes ($\omega_s$ and $\omega_i$, respectively) forming doubly resonant pairs A, B, and C. The right panel shows the gain curve narrowed by double-pass pumping and type-II phase-matching so that only one doubly resonant pair B (tuned to the peak gain) is under the gain curve. (b) Experimental setup of the bright subnatural-linewidth single-photon source. The pump laser is frequency-doubled from an external cavity diode laser (ECDL) using second-harmonic generation (SHG) in a cavity. The orthogonally polarized signal and idler photons are separated by a polarizing beam splitter (PBS) and detected by fiber-coupled single-photon-counting modules (SPCM). The beam splitter (BS) is used for the Hanbury Brown Twiss measurement. Mirrors are labeled by M. The electro-optic modulator is labeled by EOM. (c) The time-domain wave packet of the single photons. The red curve is a fitting to the experimental data (black) using Eq.~(\ref{eq:G2}). The anticorrelation parameter $\alpha_{2d}$ is shown on the right vertical axis. The horizontal dashed line represents $\alpha_{2d}=1$. The inset shows the anticorrelation parameter $\alpha_{3d}$ for several pump powers.}
\end{figure}

Our experimental setup is illustrated in Fig.~\ref{fig:1}(b). The resonant parametric down-conversion is realized in a monolithic periodically poled KTiOPO$_4$ (PPKTP) crystal, which supports type-II phase matching. The crystal is optically pumped by a continuous-wave, frequency-doubled 397.5-nm laser, of which the fundamental beam is frequency-stabilized to the rubidium D1 line using the saturated absorption spectroscopy. The temperature of the crystal is stabilized by a thermoelectric cooler with a stability better than 0.5~mK. To implement the monolithic cavity for the signal and idler photons, the end faces of the crystal are spherically polished and high-reflection coated at the signal and idler wavelengths. The end face farther away from the pump is also high-reflection coated at the pump wavelength to implement the double-pass pumping. 

To generate a single photon, a single photon counting module (350-ps timing resolution) is used to detect an idler photon and herald a single signal photon. A typical time-domain wave packet of the single photon, which is obtained by the photon correlation measurement of the signal and idler photons, is shown in Fig.~\ref{fig:1}(c). The waveform of the wave packet is described by
\begin{equation}
\label{eq:G2}
G^{(2)}(\tau) = R^2 + \frac{4 \kappa^2 \Gamma_s \Gamma_i}{(\Gamma_s + \Gamma_i)^2} \times
\left\{ \begin{array}{ll}
 e^{\Gamma_s \tau} &\mbox{, $\tau<0$ } \\
 e^{-\Gamma_i \tau} &\mbox{, $\tau>0$ }
       \end{array}, \right.
\label{eq:G2}
\end{equation}
where $\tau$ is the time delay between the detection of the signal and idler photons, $R$ is the pair rate, $\kappa$ is the coupling strength of the parametric interaction, and $\Gamma_s$ = 1/(20.7 ns) and $\Gamma_i$ = 1/(24.4 ns) are the total cavity decay rates of the signal and idler photons, respectively. The difference between $\Gamma_s$ and $\Gamma_i$ is probably due to the polarization dependence of the reflectivity of the crystal end faces and the group indices of the signal and idler photons. The bandwidth of the single photons \cite{Chuu11, Chuu12} is given by $\Delta \omega = [(\sqrt{\Gamma_s^4+6 \Gamma^2_s \Gamma^2_i + \Gamma_i^4}-\Gamma^2_s-\Gamma^2_i)/2]^{1/2} = 4.5$~MHz and is narrower than the natural linewidth of the $^{87}$Rb D1 line. The detected pair rate is $R =$ 2868~s$^{-1}$ at the pump power of 30~$\mu$W and is stable over several hours with a fluctuation of $\pm 6\%$ on the scale of seconds. Correcting for the quantum efficiency of each detector (63\%) and the transmittance in the signal and idler channel (27\% and 54\%, respectively), the generated spectral brightness is $3.67 \times 10^5$~s$^{-1}$~mW$^{-1}$~MHz$^{-1}$.

On the right vertical axis of Fig.~\ref{fig:1}(c), we show the ``two-detector" anticorrelation parameter $\alpha_{2d} = R/(\tau_c R_s R_i)$ \cite{Grangier86,Pearson10} at different time delays, where $\tau_c$ is the coincidence window of the photon correlation measurement, $R_s$ is the signal rate, and $R_i$ is the idler rate. The fact that $\alpha_{2d}$ is always larger than 1 confirms the correlation between the signal and idler photons. With a beam splitter and another single-photon counting module (D$_{s'}$) in the signal channel [Fig.~\ref{fig:1}(b)], we also measure the ``three-detector" anticorrelation parameter $\alpha_{3d} = R_{ss'i} R_{i}/(R_{si} R_{s'i})$ \cite{Grangier86,Pearson10} of the heralded single photons conditional on the detection of the idler photons. Here, $R_{ss'i}$, $R_{si}$, and $R_{s'i}$ are the coincidence rates between all three detectors, detectors D$_s$ and D$_i$, and detectors D$_{s'}$ and D$_i$, respectively. The measured $\alpha_{3d}$ at several pump powers [inset of Fig.~\ref{fig:1}(c)] are less than 0.5, indicating the single-photon nature of the heralded signal photons. 

\subsection{A. Waveform manipulation}

\begin{figure}[t]
\centering
\includegraphics[width=0.8 \linewidth]{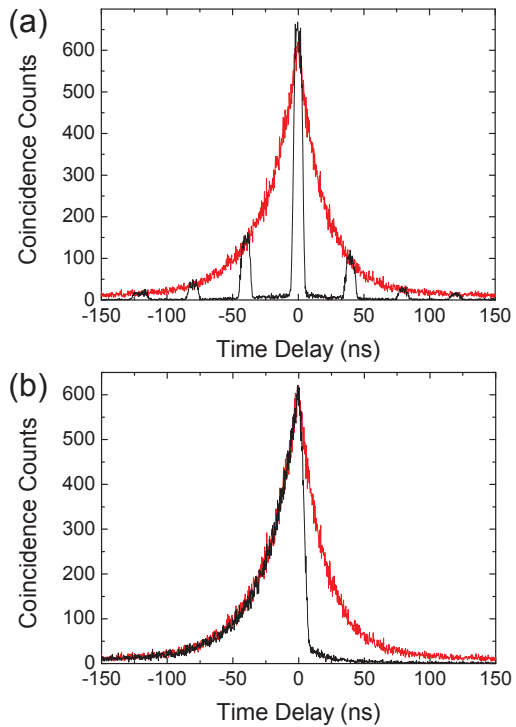}
\caption{\label{fig:2} Electro-optic modulation of the single photons. The single photons are modulated by an electro-optic modulator, which results in waveforms of (a) seven square pulses and (b) an exponential growth. The red and black curves are the waveforms of the single photons without and with modulation, respectively.}
\end{figure}

The detection of the idler photon provides a time reference for manipulating the waveform of a single signal photon. Such possibility enriches the photonic quantum technologies \cite{Cirac97, Gorshkov07, Novikova07, Zhang12, Inoue02, Liu13}. To control the waveform of our single photons, an electro-optic modulator is used to modulate the probability amplitude of the single photons. Importantly, the arrival of the single photon at the modulator is synchronized with the start of the amplitude modulation, which is triggered by the detection of the idler photon. Fig.~\ref{fig:2} shows two examples of the waveform-controlled single photons. The amplitude modulation has the shape of periodic square pulses and a Heaviside step in Figs.~\ref{fig:2}(a) and \ref{fig:2}(b), respectively, resulting in single photons with waveforms of seven square pulses and exponential growth. Previously the electro-optic modulation of single photons was demonstrated with an atomic ensemble \cite{Kolchin08}. Spectral shearing and bandwidth manipulation of heralded single photons were achieved in spontaneous parametric down-conversion \cite{Wright17, Karpinski17}.

\subsection{B. Controlled light-matter interaction}

\begin{figure}[t]
\centering
\includegraphics[width=0.8 \linewidth]{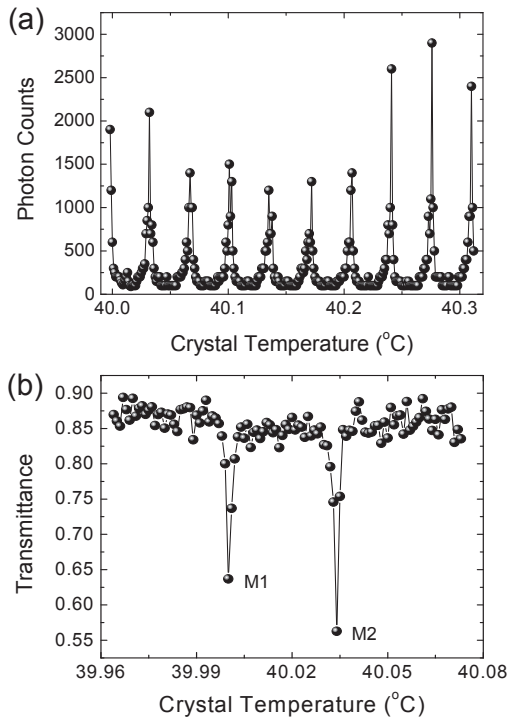}
\caption{\label{fig:3} (a) Temperature-dependent mode structure of the single photons. (b) Transmittance of a single photon through a rubidium vapor cell as a function of the crystal temperature. Modes M1 and M2 are within the Doppler broadened absorption dips. The baseline transmittance of $85~\%$ is due to the noise photons, which are reflected by the entrance and exit surfaces of the vapor cell.}
\end{figure}

We next study the controlled interaction between the single photons and atoms. This can be achieved by modifying the properties of either the single photons or the atoms. The frequency of the single photons can be tuned by changing the temperature of the monolithic PPKTP crystal. Suppose the signal and idler photons are generated in a doubly resonant mode pair. If the temperature of the crystal is changed by an amount corresponding to half the differential free spectral range (FSR), $\Delta \nu /2$=$(\Delta \nu_s-\Delta \nu_i)/2$~$\approx 220$~MHz, the signal and idler photons will be more favorably generated in the adjacent mode pair, which has a frequency difference from the previous mode by one FSR. The behavior of this temperature-dependent mode structure can be seen in Fig.~\ref{fig:3}(a), where the coincidence rate (including the accidental coincidence) between the signal and idler channels is measured for various crystal temperatures with steps of 1 mK. The strength of each doubly resonant mode depends on the magnitude of the gain. The spacing of the adjacent mode pairs is $\Delta T_m = \Delta \nu / 2 \alpha_T$~$\approx 25$~mK, where $\alpha_T = 7.8$~GHz/K is the temperature tuning coefficient. 

Fig.~\ref{fig:3}(b) shows the transmittance of the single photons through a $^{87}$Rb vapor cell controlled by tuning the frequency of the single photons. For this purpose, we lock the pump frequency to the $\left| 5{\rm S}_{1/2}, F = 3 \right\rangle \rightarrow \left| 5{\rm P}_{1/2}, F = 2 \right\rangle$ transition of $^{85}$Rb. This would allow us to observe the transmission dips at two Doppler-broadened transitions $\left| 5{\rm S}_{1/2}, F = 2 \right\rangle \rightarrow \left| 5{\rm P}_{1/2}, F = 1 \right\rangle$ and $\left| 5{\rm S}_{1/2}, F = 1 \right\rangle \rightarrow \left| 5{\rm P}_{1/2}, F = 2 \right\rangle$ of $^{87}$Rb, which have a frequency difference close to the mode spacing of our single photons. As shown in Fig.~\ref{fig:3}(b), the transmission dips are observed when the single photons are generated in adjacent modes M1 and M2, of which the frequencies fall in the Doppler-broadened transitions.

\begin{figure}[t]
\centering
\includegraphics[width=0.8 \linewidth]{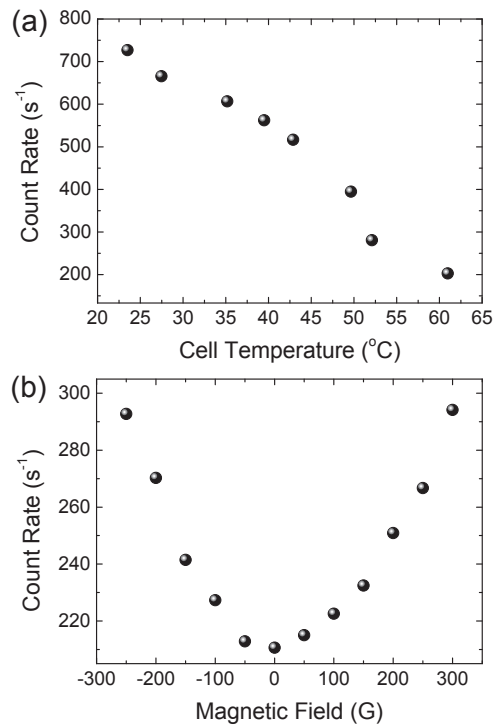}
\caption{\label{fig:4} Absorption of the single photons controlled by changing (a) the cell temperature and (b) the external magnetic field. The change of the cell temperature (external magnetic field) corresponds to the modification of the atom density (resonance frequency). The magnitudes of the magnetic fields in (b) are measured at the center of the cell. The non-uniformity of the magnetic field across the cell is less than 10\%.}
\end{figure}

The transmittance of the single photons can also be controlled by modifying the properties of the rubidium atoms. In Fig.~\ref{fig:4}(a), the transmittance of the single photons in mode M1 is shown for various temperature of the vapor cell. When the temperature increases, the density of the rubidium atoms increases. As a result, the absorption of the single photons by the atoms increases and the transmittance reduces accordingly. 

In Fig.~\ref{fig:4}(b), a dc magnetic field with various magnitudes is applied across the vapor cell while we measure the count rate of the transmitted single photons. The magnetic field, created by a solenoid, introduces a Zeeman shift on the resonance frequency of the rubidium atoms. As the magnetic field increases (in either direction), the single photons is detuned away from the resonance and the transmission of the single photons thus increases. The maximum absorption occurs near zero magnetic field, which indicates that the frequency of the single photons in mode M1 is close to the resonance of the rubidium atoms. Fine tuning of the single photon frequency may be possible by changing the pump frequency.

\section{IV. Conclusion}

In summary, we report a bright single-photon source with a subnatural linewidth of 4.5 MHz, a generated spectral brightness of $3.67 \times 10^5$~s$^{-1}$~mW$^{-1}$~MHz$^{-1}$, and a controllable waveform. The controlled interaction between the single photons and atoms is also demonstrated by the absorption in a rubidium vapor. The near-infrared single photons are advantageous for photonic quantum technologies due to their compatibility with the telecommunication fiber technology and the availability of high-quantum-efficiency single-photon-counting modules. The miniature design of the single-photon source is also suitable for scaling up. In addition, the long temporal width of the single photons provides the opportunity for modulating the phase of the single photons. Such possibility can find applications in the manipulation of two-photon interference \cite{Specht09}, quantum key distribution protocols \cite{Inoue02, Tittel00, Marcikic01}, or single-photon hiding~\cite{Belthangady10}. Due to the probabilistic nature of spontaneous parametric down-conversion, the generation of the heralded single photons is not deterministic. However, with the technique of time multiplexing (for example, \cite{Kaneda15}), the nondeterministic nature can be overcome to increase the single-photon probability. The reflection of the double-pass pump on the crystal surface could lead to phase shift  that results in biphoton generation at a nonideal phase-mtaching temperature. To resolve this complication, one may implement the double-pass pump with an external mirror or use the electro-optic effect to tune the refractive indices.

\section{ACKNOWLEDGMENTS}

The authors thank I. A. Yu, I.-C. Chen, C.-I. Jiang and Z.-M. Peng for helpful discussion. This work was supported by the Ministry of Science and Technology, Taiwan (MOST 103-2112-M-007-015-MY3).




\end{document}